\documentclass[aps,prd,floatfix,superscriptaddress,preprintnumbers,nofootinbib]{revtex4-1}
\usepackage{psfrag,graphicx,epsfig,amsmath,amssymb,bm,tikz,multirow,booktabs}
\usetikzlibrary{arrows,decorations.markings,decorations.pathmorphing}
\usepackage[font=footnotesize,justification=raggedright]{caption}

\expandafter\let\csname equation*\endcsname\relax
\expandafter\let\csname endequation*\endcsname\relax
\usepackage{amsmath}
\usepackage{graphicx}

\def\unam{Universidad Nacional Aut\'onoma de M\'exico, Distrito Federal, M\'exico}

\def\cabib{Centro At\'omico Bariloche - Instituto Balseiro, CNEA/CONICET, Argentina}
\def\fnal{Fermi National Accelerator Laboratory, Batavia, IL, United States}
\def\ufdrj{Universidade Federal do Rio de Janeiro, Instituto de F\'isica, Rio de Janeiro, Brazil}
\def\zurich{Universit\"at Z\"urich Physik Institut, Zurich, Switzerland}

\def\fiuna{Facultad de Ingenier\'ia - Universidad Nacional de Asunci\'on, Paraguay}

\def\cbpf{Centro Brasileiro de Pesquisas Fisicas, Rio de Janeiro, Brazil}
\def\unsa{Depto. de Ing. Electrica y de Computadores, Universidad Nacional del Sur, Bahia Blanca, Argentina}
\def\unsb{Comisi\'on de Investigaciones Cient\'ificas Provincia Buenos Aires, La Plata, Argentina.}
\def\unsc{Instituto de Investigaciones en Ing. Eléctrica "Alfredo Desages", CONICET - Universidad Nacional del Sur, Bahía Blanca, Argentina.}
\def\unsd{Depto. de Ingeniería, Universidad Nacional del Sur, Bahía Blanca, Argentina.}
\def\pucrj{Pontificia Universidade Católica, Rio de Janeiro, Brazil}
\begin{document}
%
\title[]{Results of the engineering run of the Coherent Neutrino Nucleus Interaction Experiment (CONNIE)}

\author{A.~Aguilar-Arevalo} \address{\unam}
\author{X.~Bertou} \address{\cabib}
\author{C.~Bonifazi} \address{\ufdrj}
\author{M.~Butner} \address{\fnal}
\author{G.~Cancelo} \address{\fnal}
\author{A.~Casta\~neda~V\'azquez} \address{\unam}
\author{C.R.~Chavez} \address{\fiuna}
\author{H.~Da Motta} \address{\cbpf}
\author{J.C.~D'Olivo} \address{\unam}
\author{J.~Dos Anjos} \address{\cbpf}
\author{J.~Estrada} \address{\fnal}
\author{G.~Fernandez~Moroni} \address{\unsa} \address{\unsc}
\author{R.~Ford} \address{\fnal}
\author{A.~Foguel} \address{\ufdrj}\address{\cbpf}
\author{K.P.~Hern\'andez~Torres} \address{\unam}
\author{F.~Izraelevitch} \address{\fnal}
\author{H.P.~Lima Jr.} \address{\cbpf}
\author{B.~Kilminster} \address{\zurich}
\author{K.~Kuk} \address{\fnal}
\author{M.~Makler} \address{\cbpf}
\author{J.~Molina} \address{\fiuna}
\author{G.~Moreno-Granados} \address{\unam}
\author{J.M.~Moro} \address{\unsd}
\author{E.E.~Paolini} \address{\unsa}\address{\unsb}
\author{M.~Sofo~Haro} \address{\cabib}
\author{J.~Tiffenberg} \address{\fnal}
\author{F.~Trillaud} \address{\unam}
\author{S.~Wagner} \address{\cbpf}\address{\pucrj}

\begin{abstract}
The CONNIE detector prototype is operating at a distance of 30 m from the core of a 3.8 GW$_{\rm th}$ nuclear reactor with the goal of
establishing Charge-Coupled Devices (CCD) as a new technology for the detection of coherent elastic neutrino-nucleus scattering. We report on the results
of the engineering run with an active mass of 4 g of silicon. The CCD array is described, and the performance observed
during the first year is discussed. A compact passive shield was deployed for the detector, producing an
order of magnitude reduction in the background rate. The remaining 
background observed during the run was stable, and dominated by internal contamination in the detector packaging materials.
The {\it in-situ} calibration of the detector using  X-ray lines from fluorescence demonstrates good stability of the readout system.
The event rates with the reactor on and off are compared, and no excess is observed coming from 
nuclear fission at the power plant. The upper limit
for the neutrino event rate is set two orders of magnitude above the expectations for the standard
model. The results demonstrate the cryogenic CCD-based detector can be remotely operated
at the reactor site with stable noise below 2 e$^-$ RMS and stable background rates.  The success of the engineering test
provides a clear path for the upgraded 100 g detector to be deployed during 2016.
\end{abstract}

\pacs{1315, 9440T}

\maketitle

\section{Introduction}


After the discovery of neutral current neutrino interactions in 1973 by Hasert et al. \cite{Hasert 1973}, it was pointed out that coherent enhancement of the elastic scattering cross section should occur \cite{Freedman 1974}. Unfortunately, even with coherent enhancement, the Coherent Elastic Neutrino-Nucleus Scattering (CENNS) has been impossible to detect  because of its very low cross section ($< 10^{-39}$ cm$^2$) \cite{Freedman 1977} and the very low energy deposition, below ~15 keV for most detector targets. Up to recent years, detector technology has not been able to provide the suitable detector mass and very low energy threshold needed for this observation. Nevertheless, the interest in low energy neutrino physics has been growing, motivated mostly by verification of the SM prediction and potential for new physics \cite{Scholberg 2006}. In astrophysics, the understanding of MeV-neutrinos is relevant for energy transport in supernovas and is a limiting factor in ongoing efforts for developing new supernova models \cite{Wilson 1974, Horowitz 2004}. Moreover, in recent years there has been a growing interest in nuclear reactor monitoring using neutrinos \cite{Barbeau 2003,Hagmann 2004}.


The coherent scattering from solar, atmospheric and diffuse supernova neutrino background has been identified as a limiting background for future dark matter Searches \cite{Snowmass 2013}. Ultimately, direct-detection experiments will start to see signals from coherent scattering of these neutrinos. Although interesting in their own right, these neutrino signals will require background subtraction or directional capability in WIMP direct detection experiments to separate them from possible dark matter signals (see  Refs. \cite{Billard 2014}, \cite{Strigari 2009} and \cite{Glutein 2010}  for earlier studies about the neutrino background in DM experiments). However, as noted above, CENNS has never been measured. The expected rates of coherent events in Ref. \cite{Snowmass 2013}
are based on a prediction in the Standard Model that has not been confirmed. There are several extensions of the SM that would result in a significant enhancement of the cross section and an increase in the rate of events by several orders of magnitude \cite{Harnik 2012}.  Enhancements to the event rates have been predicted in various models including those in which a neutrino has a magnetic moment, those in which neutrino-nucleus scattering is mediated by a light boson, and those in which 2\% of the solar neutrino flux oscillates into a Standard Model singlet $\nu_{\mbox{s}}$ which couples to atomic nuclei via a light U(1)B gauge boson. The detection of coherent elastic scattering will also open a new window for the study of neutrino
oscillations and the search for sterile neutrinos \cite{Anderson 2012} \cite{Dutta 2015}.


Since the discovery in 1956, the prefered technology for the detection of reactor neutrinos has been the inverse beta decay on free protons
(threshold 1.8 MeV). It has been exploited by many experiments with an emphasis in the study of neutrino oscillations \cite{doublechooz, Boehm 2000, kamland, Zacek, Vidyakin, Declais, An, Ahn, Abe}.
These experiments use large volume of target material to balance their high thresholds
( $\sim$10 keV). In recent years, with the decreasing threshold of solid-state detectors, there has been a growing interest in using them as neutrino detectors \cite{Xin 2005, GEMMA} \cite{Wong 2007}.  The Coherent Collaboration \cite{Coherent 2015} is using multiple detector technologies to attempt to detect coherent scattering from neutrinos produced in muons decaying at rest at the Spallation Neutron Source (SNS). The neutrinos produced at SNS have higher energies than reactor neutrinos, and therefore produce larger recoil signals accessible to higher threshold detectors. However, the new physics discussed in Ref. \cite{Harnik 2012} will manifest only at low energies, and will be harder to observe using neutrinos from muon decays.  On the other hand, improvements in CCD technology has allowed the development 
of devices with larger mass, which together with their very low energy threshold and good  spatial resolution, makes them a viable 
option to detect CENNS.

This paper is organized as follows: in Sec. II the detector concept, the prototype and the installation at the Angra II reactor are described; Sec. III discuses the prototype performance during the engineering run and Sec. IV compares the measurements with the reactor ON and OFF. The conclusions
and future prospects are presented in Sec. V.

\section{The CONNIE Detector at Angra-II}

\begin{figure}[t]
\begin{center}
\includegraphics[width=0.8\textwidth]{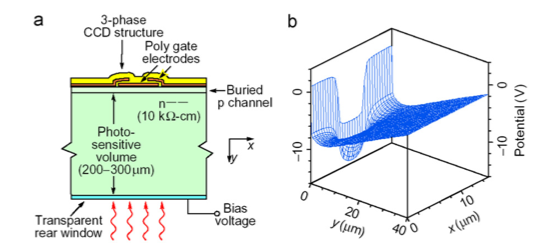}
\vspace{-10pt}
\caption{Pixel cross section of a 250 um thick CCD developed on Lawrence Berkeley Laboratory. The electrostatic potential generated by the three phases under the gates is shown as a function of depth (y-axis) and one of the lateral directions (x-axis). Figure from \cite{Oluseyo 2004}.} 
\label{fig:CCDlayout}
\end{center}
\end{figure}

The Coherent Neutrino Nucleus Interaction Experiment (CONNIE) is a solid-state based detector installed at a nuclear
power plant. At the core of the CONNIE detector there is an array of Charge-Coupled Devices (CCDs).
Although originally devised as memory devices \cite{Boyle 2010, Smith 2010}, CCDs have found a niche as imaging detectors due to their 
ability to obtain high-resolution digital images. In particular, scientific CCDs have been used extensively 
in ground and space-based astronomy and X-ray imaging \cite{Janesick 2001}. CCDs have high detection efficiency, low noise, 
good spatial resolution, and low dark current. Thick CCDs with increased active mass have also
been used as particle detectors in the DAMIC search for dark matter \cite{Holland 2003, Barreto 2012}. 
A sketch of a CONNIE CCD is shown in Fig. \ref{fig:CCDlayout}. 

\begin{figure}[t]
\begin{center}
\includegraphics[width=0.6\textwidth]{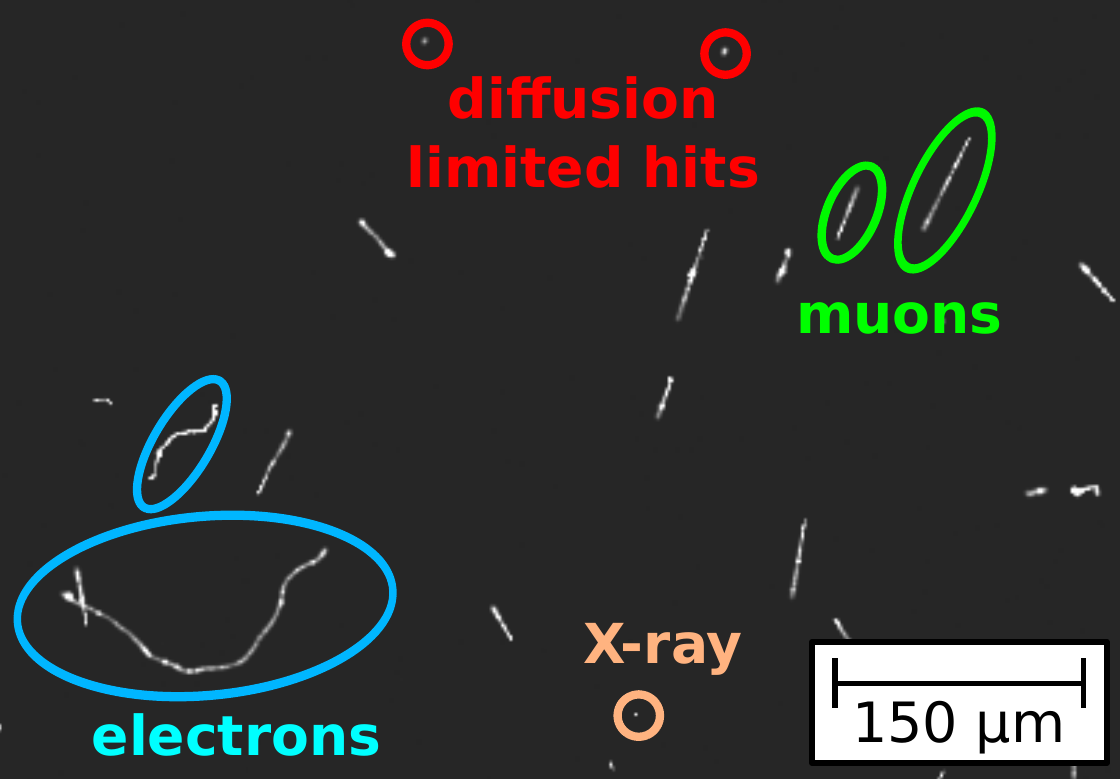}
\vspace{-10pt}
\caption{Sample of tracks produced by different types of radiation in a CCD. Muons and Compton electrons can be clearly
distinguished from X-ray and other diffusion-limited hits. For diffusion-limited hits, the size of the reconstructed charge cloud is
determined by the spread of the charge inside the silicon, and not by the range of the ionizing particles. The nuclear
recoil produced from a coherent scattering with a neutrino will result in a diffusion-limited hit.} 
\label{fig:partid}
\end{center}
\end{figure}

\begin{figure}[t]
\begin{center}
\includegraphics[width=0.5\textwidth]{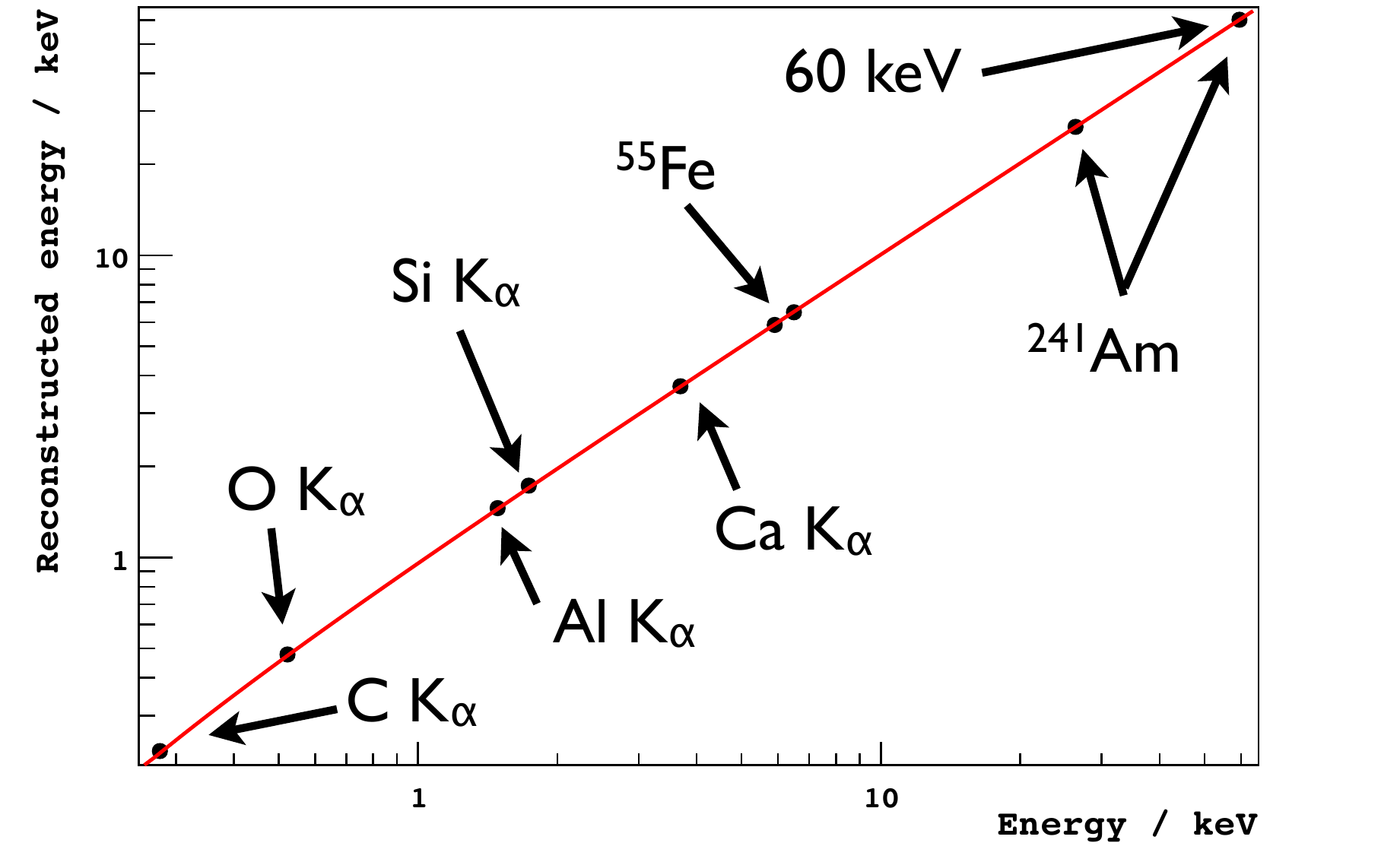}\includegraphics[width=0.5\textwidth]{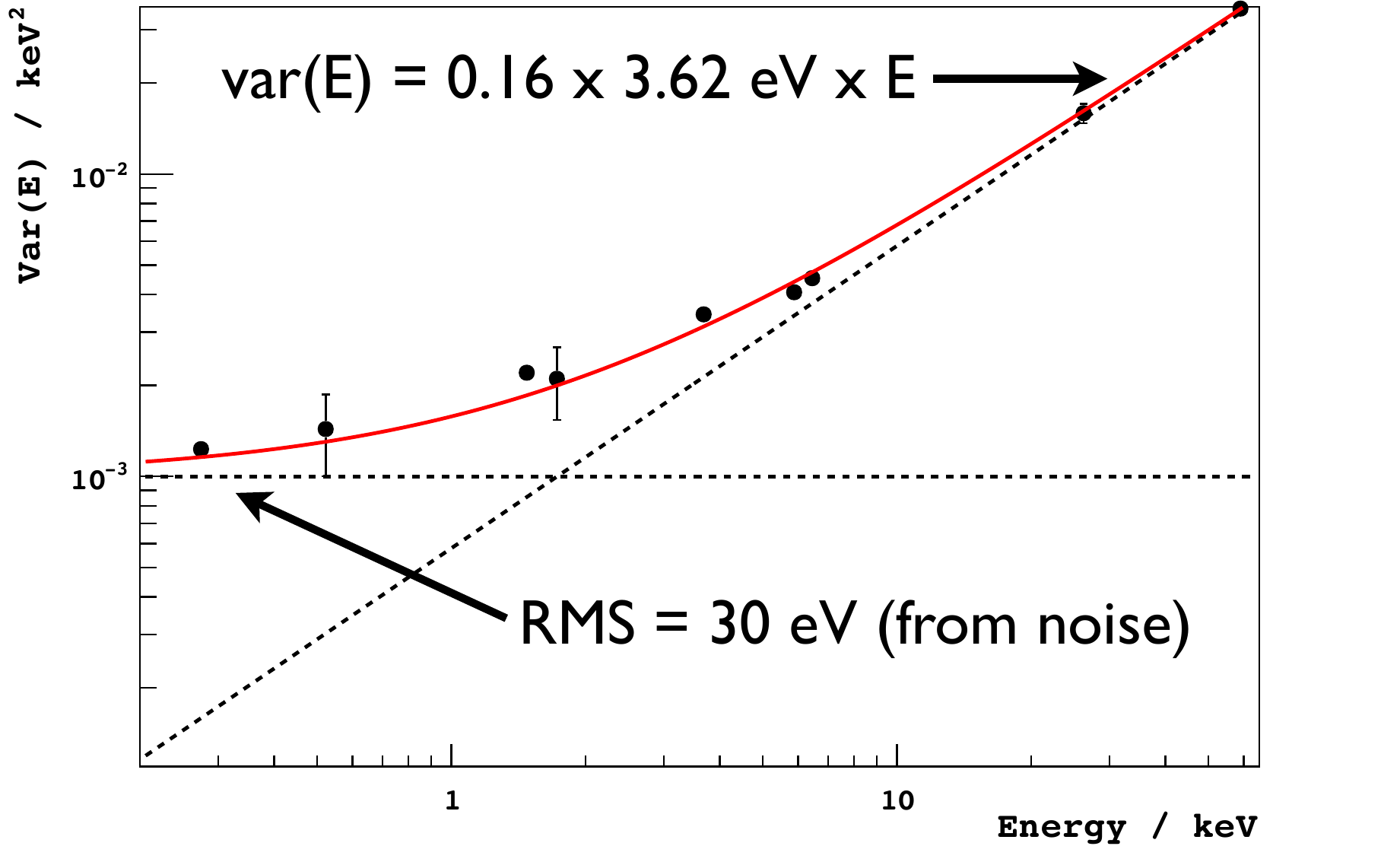}
\vspace{-10pt}
\caption{Left: Linearity of CCD response to several X-ray lines from 277 eV (C fluorescence) to 60 keV ($^{241}$Am source). Right: Energy
resolution for X-ray lines reconstructed in the CCD.} 
\label{fig:ccdcal}
\end{center}
\end{figure}

The performance of thick CCDs for particle
identification has been discussed extensively in Refs. \cite{Tiffenberg 2013} and \cite{Aguilar 2015}.
The expected signature from neutrino-nucleus coherent scattering is a diffusion-limited hit
as shown in Fig. \ref{fig:partid}. The energy calibration of CCDs for electronic
recoils has been studied in detail in Ref. \cite{Tiffenberg 2013}, and a summary is shown in Fig. \ref{fig:ccdcal}. 
For nuclear recoils some fraction of the energy does not produce ionization through the so-called quenching process. The calibration
of low-energy nuclear recoils in silicon is thus more challenging, and not
as well measured. This calibration is currently under study \cite{Antonella_cal,Uchicago_cal}, but for the purpose of this
work we will consider the Lindhard quenching model discussed in Refs. \cite{Lindhard 1963} and \cite{Lewin 1996}.

\begin{figure}[t]
\begin{center}
\includegraphics[width=0.46\textwidth]{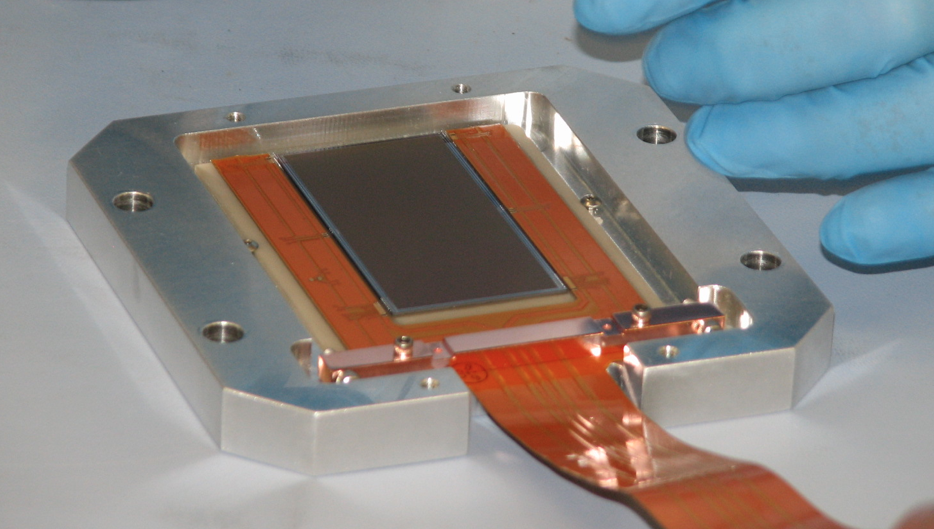}
\includegraphics[width=0.2\textwidth]{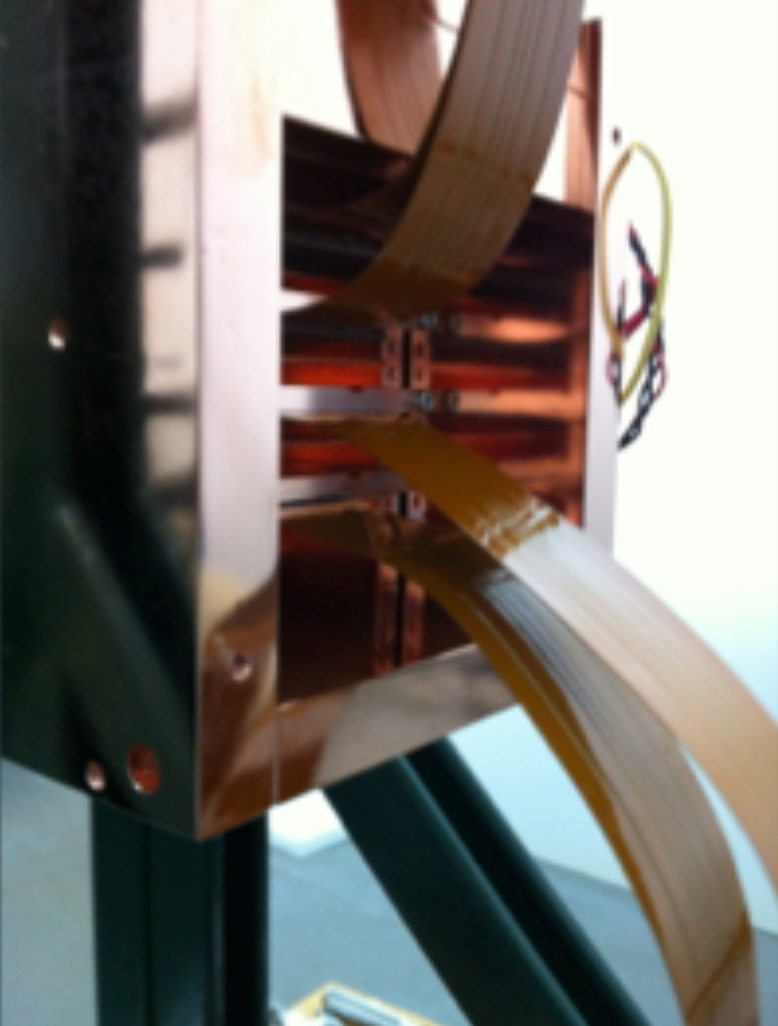}
\vspace{-0pt}
\caption{Left: CONNIE CCD engineering detector package shown in an aluminum carrier used for handling the detectors in the lab. Right:
Cold box used for detector array at CONNIE with 4 sensors in place.} 
\label{fig:boxphoto}
\end{center}
\end{figure}

The CCD detectors used for CONNIE are operated at -140$^{\circ}$ C to reduce the thermal dark current generated in
the silicon. This is achieved inside a vacuum vessel ($10^{-7} \mbox{torr}$) and cooled with a Gifford-McMahon close-cycle
refrigerator \cite{cryomech}. 
The temperature is controlled to better than 0.1$^{\circ}$ C using a commercial PID system \cite{lakeshore}. The CCDs are packaged inside
a copper box acting as a radiation shield. The box is kept at -140$^{\circ}$ C, which also reduces 
the infrared radiation reaching the active surface of the CCDs (see Fig. \ref{fig:boxphoto}).

\begin{figure}[t]
\begin{center}
\includegraphics[width=0.5\columnwidth]{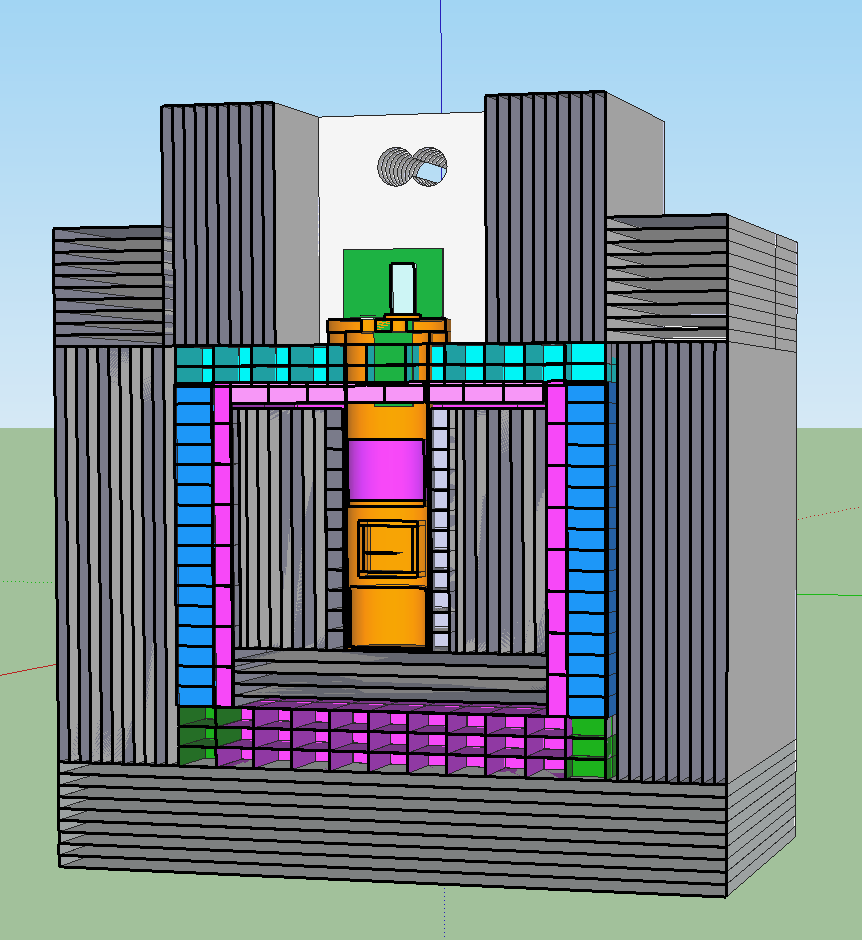}
\caption{\footnotesize Radiation shield for the CONNIE experiment. The shield has an inner layer of 30 cm polyethylene (grey), followed by 15 cm of lead (blue, pink and green), and an additional outer layer of 30 cm of polyethylene (grey). The CCD detector array is inside the copper vessel shown in orange.} 
\label{fig:shield}
\end{center}
\end{figure}

The CONNIE experiment is located in a temporary laboratory space, built inside
a standard shipping container located outside of the main reactor building. The experiment has no overburden
for radiation shielding. The vacuum vessel is installed inside the radiation shield as shown in Fig. \ref{fig:shield}. The shield consists of an 
inner layer of 30 cm of polyethylene, followed by 15 cm of lead, and an additional 
outer layer of 30 cm of polyethylene. Lead is a good shield for gammas, while polyethylene
is an efficient shield for neutrons. Since neutrons are produced when cosmic muons interact with lead,
a fraction of the polyethylene shield is kept inside the lead layer. There is also a 15 cm lead cylinder inside
the vacuum vessel, above the cold box containing the detectors. This cylinder shield the
CCDs from radioactive components in the readout electronics.

The detector is located at a distance of 30 m from the core of the Angra-2 nuclear reactor in the 
Almirante Alvaro Alberto Nuclear Power Plant, in the state of Rio de Janeiro, Brazil. Angra-2 is a Pressurized Water Reactor (PWR)
with a thermal power of 3764 MW that started commercial operation during the year 2000,  the historical operation factor
for the facility is 88\%  \cite{iaea_pris}. In steady-state operation, the total neutrino flux produced by the reactor is $1.21 \times 10^{20}$ $\bar{\nu}/$s , 
and the flux density at the detector (L = 30 meters from the core) is $7.8 \times 10^{12}$ $\bar{\nu}/$cm$^2/$s. The large flux 
justifies the use of this nuclear reactor as a neutrino source for the CONNIE experiment \cite{Moroni 2015}.


\section{Performance of the Prototype CCD Array}


The prototype CCD array installed at the Angra-2 power plant has four CCD sensors. Each sensor has an active mass of 1 gram, 
area of 18 cm$^2$ (6 cm $\times$ 3 cm), and is 250 $\mu$m thick. The sensors are divided into 8 MPix (2k $\times$ 4k), each pixel being 15 $\mu$m $\times$ 15 $\mu$m. 
The detectors used for this installation are engineering-grade sensors because of their lower cosmetic quality, having several defects
that produce imaging artifacts. Two of them (A,B) in the engineering array are considered of sufficient
quality for the data analysis presented here.

\begin{figure}[t]
\begin{center}
\includegraphics[width=0.9\textwidth]{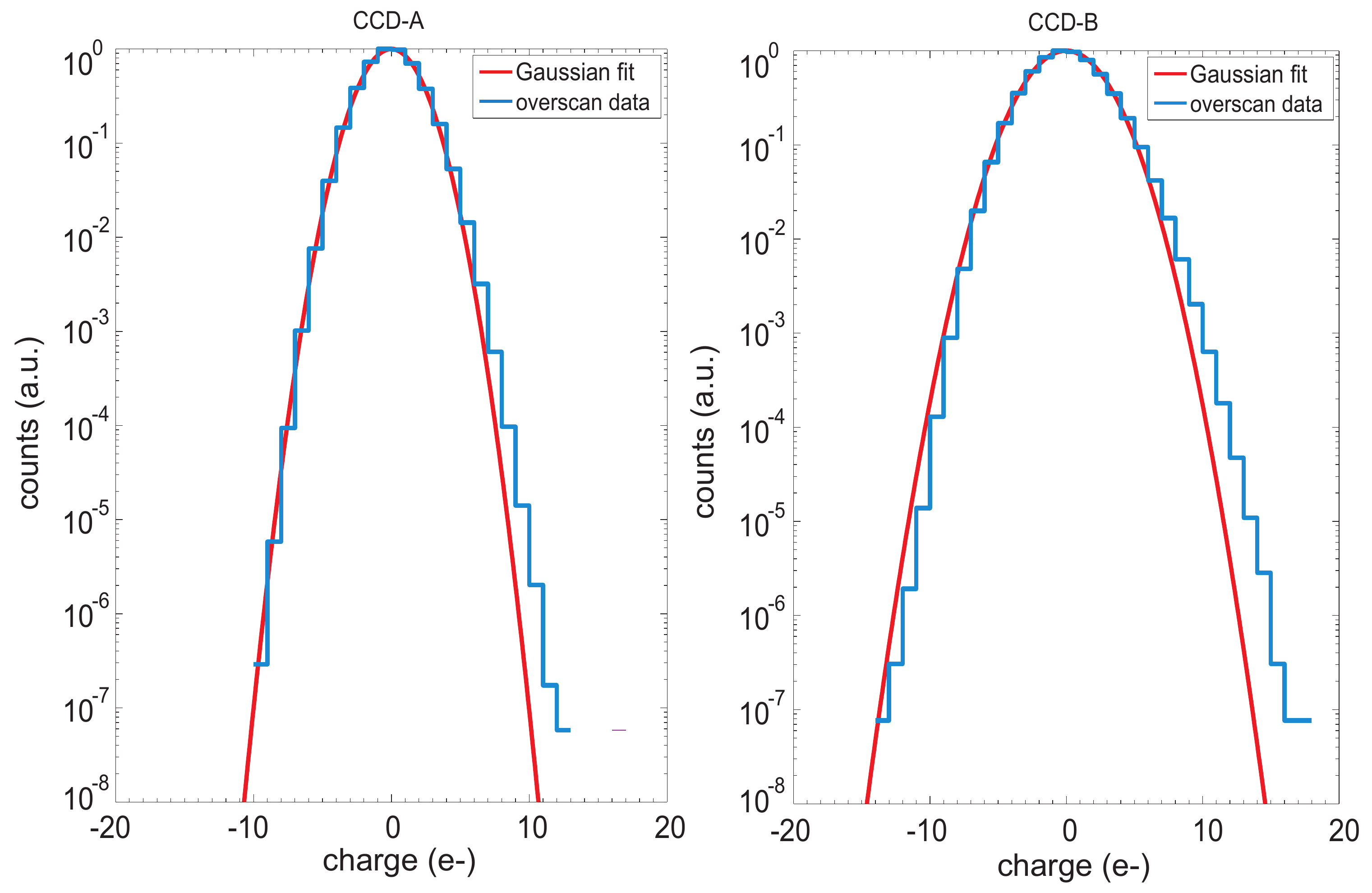}
\caption{Charge distribution for pixels without signal (overscan) in CCD-A and CCD-B. The width
of the distribution is used to measure the noise in each sensor. }
\label{fig:noise}
\end{center}
\end{figure}

The key performance parameter of the CCDs that make the CONNIE experiment possible is the 
readout noise. The noise measured on pixels with zero exposure time (so-called overscan pixels \cite{Janesick 2001}) 
for the detectors operating at the Angra-2 nuclear power plant is shown in Fig. \ref{fig:noise} . 
The noise achieved in the system is 1.8 (2.4) e$^-$ RMS for CCD-A (CCD-B), equivalent to 7 eV (9 eV) of ionization energy 
considering 3.745 eV/e in silicon at this temperature \cite{conversionFactor}. 

\begin{figure}[t]
\begin{center}
\includegraphics[width=0.8\textwidth]{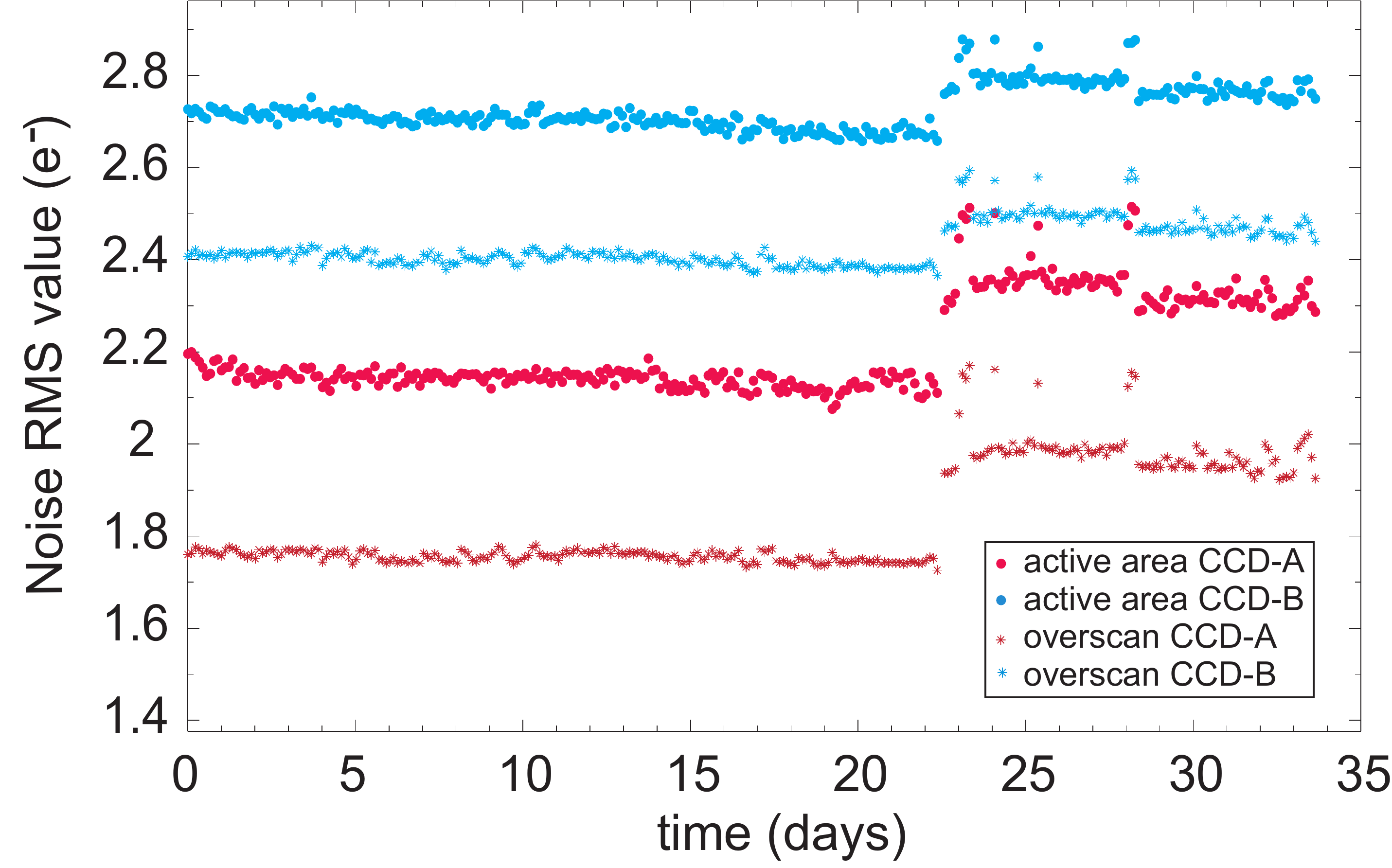}
\caption{Noise for CCD-A and CCD-B as a function of time.  The zero
along the X-axis is selected at the start of the data collection period on 12-OCT-2015.
The electrical grounding of the DAQ was changed in day 23, as discussed in the text. }
\label{fig:noisestab}
\end{center}
\end{figure}

The noise as a function of time is shown in Fig. \ref{fig:noisestab} .The plot shows two
data sets for each detector. The overscan data corresponds to the noise measured with no exposure time ($<$1 msec), generated
by continuous readout of the output stage for the sensor without any charge. The noise measured in the overscan is representative of the
readout noise of the CCD. The active area data corresponds to the noise measured in the pixels with 8700 second exposure time,
and includes the fluctuations produced by dark current. The results show higher noise in CCD-B, associated with the lower quality of the output
stage of this sensor. The dark current increases the noise on the image by about 20\%, this represents a dark current
rate of $\sim$ 0.4 e$^-$/pix/hour. This level of dark current is consistent with the expectations
for the engineering sensors, and has been reduced to $10^{-3}$ e-/pix/day in scientific grade detectors \cite{Aguilar 2015}.
The plot in Fig. \ref{fig:noisestab} demonstrates the susceptibility of the CONNIE readout system to external noise: on day 23 the 
electrical power configuration was changed to share the AC with other equipment at the reactor site, and the noise increased by
approximately 10\%.

\begin{figure}[t]
\begin{center}
\includegraphics[width=0.9\textwidth]{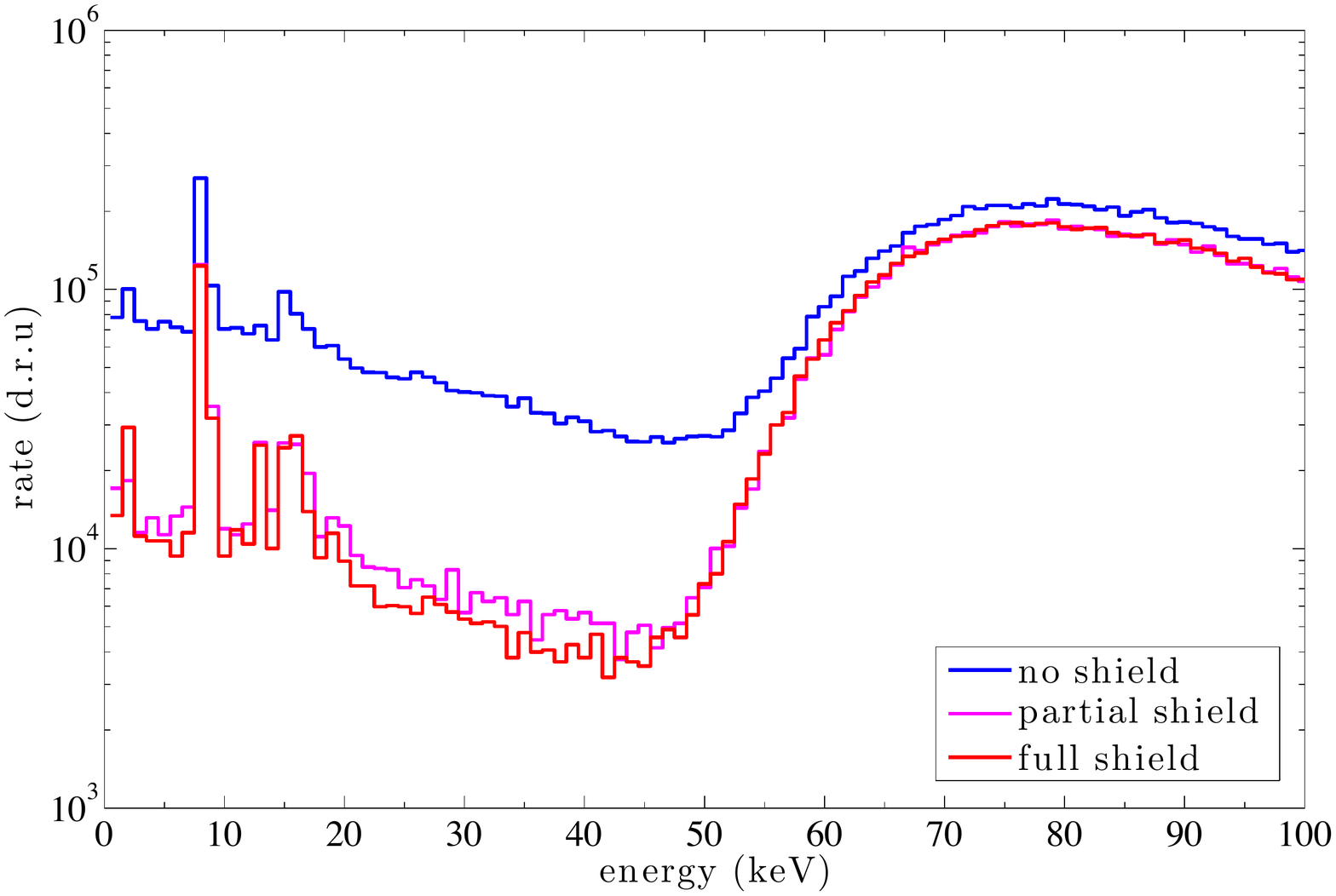}
\caption{Spectrum measured in the CONNIE engineering array with no shield (green), partial shield (red), and full shield in
events/kg/day/keV (d.r.u.). The features of this spectrum are discussed in the text.}
\label{fig:shieldper}
\end{center}
\end{figure}

The detection of the coherent scattering of neutrinos is done comparing 
data collected with reactor on (RON) and reactor off (ROFF). The radiation background is the ultimate
limit for the sensitivity of the experiment.  The event rate measured with different configurations  of the radiation shield is shown in 
Fig. \ref{fig:shieldper}. With no shield a rate of 2$\times 10^5$ events/kg/eV/day is
observed at an energy of 20 keV. Using the full shield described in Fig.\ref{fig:shield}, the rate is
reduced by approximately one order of magnitude. The
bump around 80 keV,  is characteristic of muon tracks traversing 250 $\mu$m thick silicon detectors.
As expected, the muon component of the spectrum is not efficiently shielded with 60 cm of polyethylene and 
15 cm of lead. The spectrum with
partial shield configuration in Fig.\ref{fig:shieldper} corresponds to 30 cm of polyethylene and 5 cm of lead.

\begin{figure}[t]
\begin{center}
\includegraphics[width=0.9\textwidth]{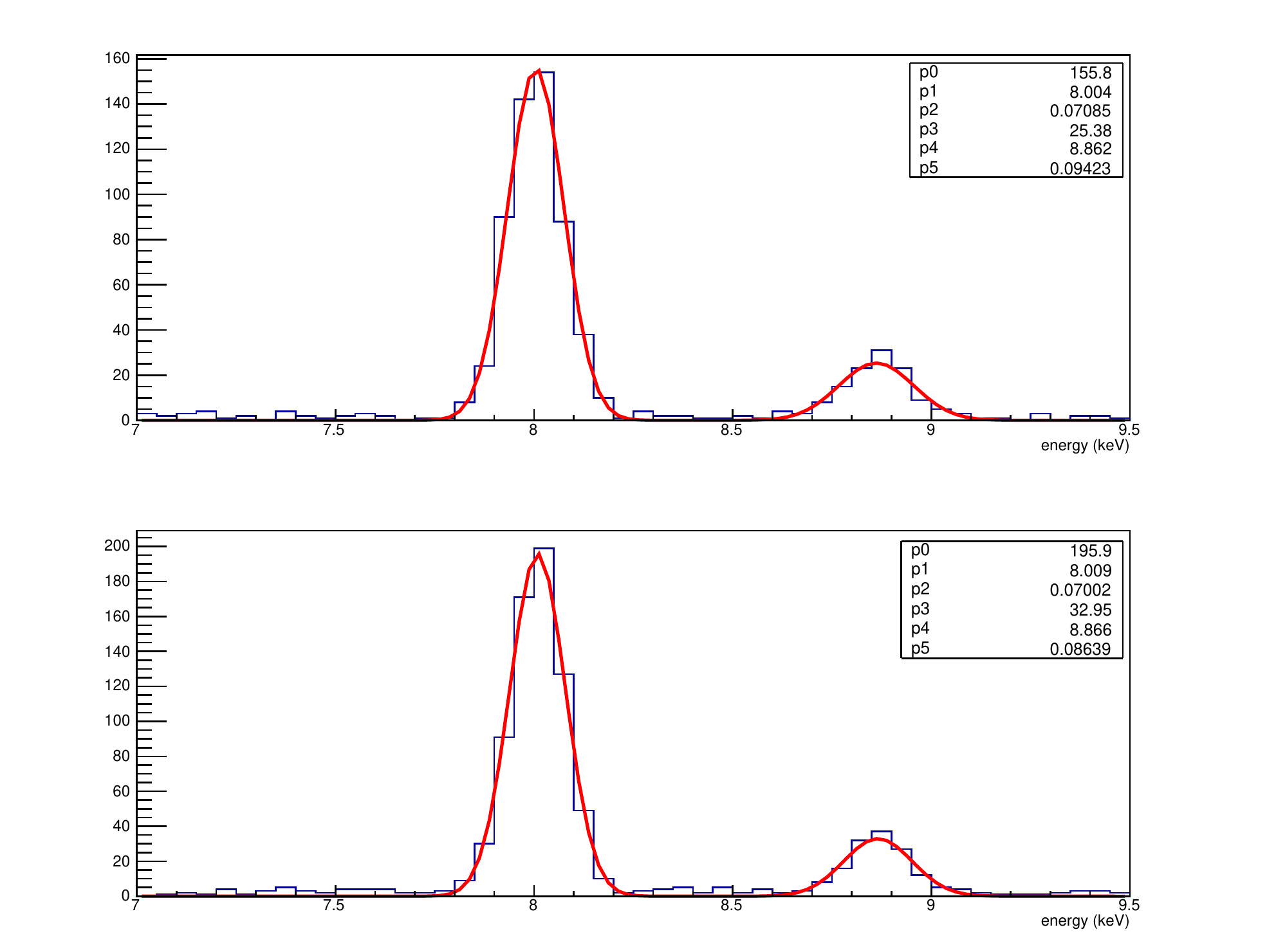}
\caption{Cu fluorescence peaks for data collected with ROFF (top) and RON (bottom) in Table \ref{tab:cupeak}.  The
fitting function in red is the sum of two gaussians for the  k$_{\alpha}$ and k$_{\beta}$ peaks
with 6 parameters: p0, p1, and p2  are the amplitude, central value and width of K$_{\alpha}$ peak, 
 p3, p4, and p5  are the amplitude, central value and width of K$_{\beta}$ peak. The results show
 good stability of the readout system, with a stable calibration and energy resolution over
 the two data collection periods.}
\label{fig:cupeak}
\end{center}
\end{figure}

\begin{figure}
\begin{center}
\includegraphics[width=.8\textwidth]{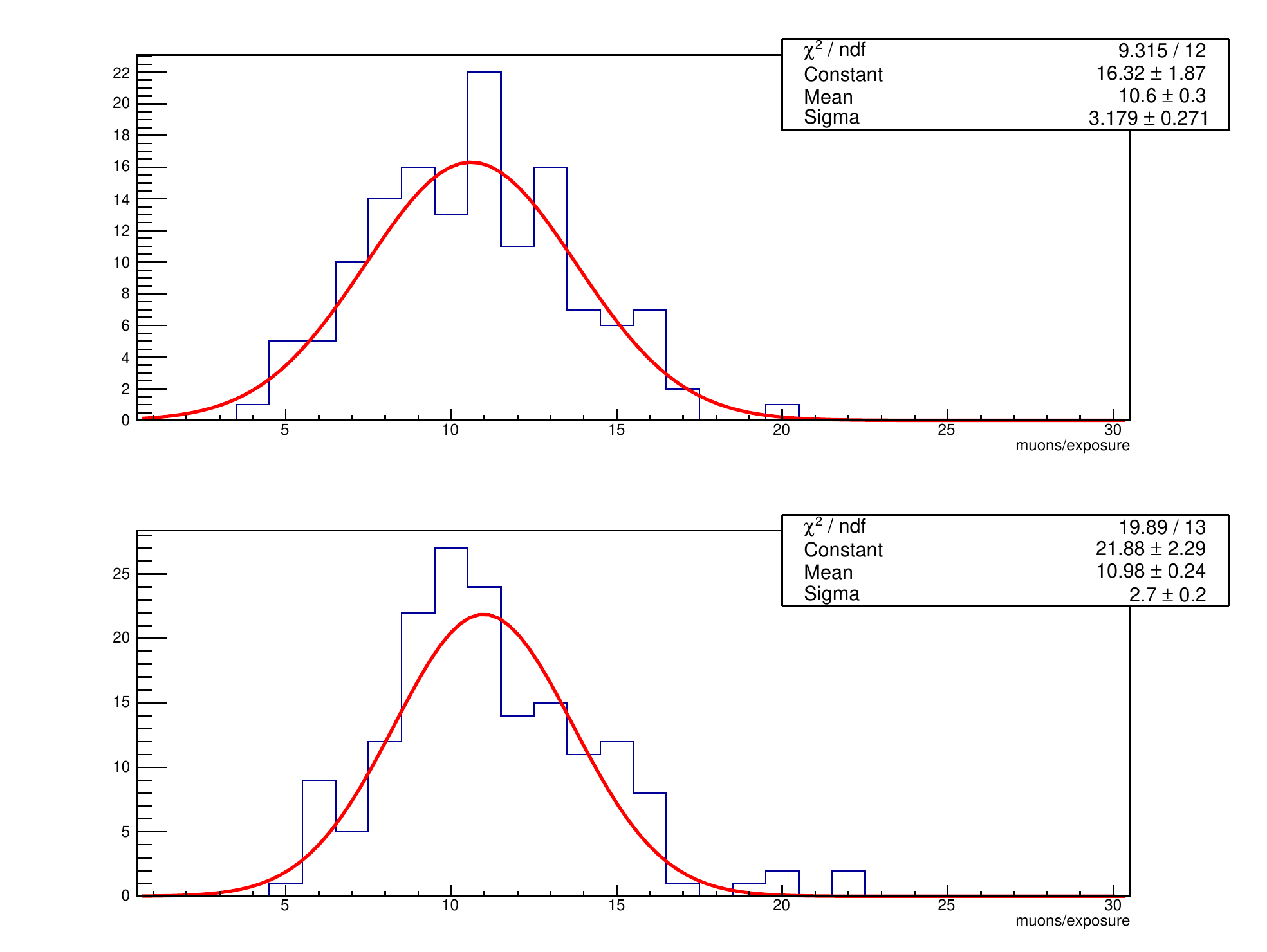}
\caption{Muon events detected for each 8700 second exposure for  ROFF (top) and RON (bottom). The rate
of  10.8 $\pm$ 0.2 muons/exposure is consistent between both periods.}
\label{fig:muonstab}
\end{center}
\end{figure}

At lower energies, the X-ray fluorescence lines produced by the detector 
materials are observed in Fig. \ref{fig:shieldper}, the most significant being those from Cu around 8 keV. 
The observed fluorescence X-ray lines for Cu are a powerful tool to monitor the 
gamma background in the detector, and to perform and {\it in-situ} calibration of
the energy scale. Figure \ref{fig:cupeak} shows the two X-ray lines produced by fluorescence in the 
Cu materials surrounding the detector (see Fig. \ref{fig:boxphoto}), these lines
are generated by the K$_{\alpha}$ and K$_{\beta}$  energy levels at 8.05 keV and 8.9 keV. The rate of events observed with RON 
and ROFF is constant as shown in Table \ref{tab:cupeak}, which indicates stability to better than 3\% of the gamma background in the experiment. This
stability is critical for the successful detection of the neutrino events. As shown in Fig. \ref{fig:partid} muons can be identified in the
images by straight tracks characteristic of minimum ionizing particles (MIPs). The rate of reconstructed muons per image during the RON and ROFF
periods is shown in Fig. \ref{fig:muonstab} to demonstrate the stability of this background component.

\begin{table}
\caption{Rates measured for the Cu fluorescence X-rays  with reactor on and reactor off}\label{tab:cupeak}
\begin{center}
  \begin{tabular}{ || c | c | c | r || }
    \hline
    Reactor &  counts            &  exposure &  rate \\
                  &   (7.8-8.2 keV) &  day &  counts/day \\ \hline \hline
    RON 	&  693				      & 18.0  &   37 $\pm$ 1 \\  \hline
    ROFF 	&  557				      & 14.8  &   38 $\pm$ 2 \\
    \hline
  \end{tabular}
\end{center}
\end{table}

\begin{figure}
\begin{center}
\includegraphics[width=0.5\textwidth]{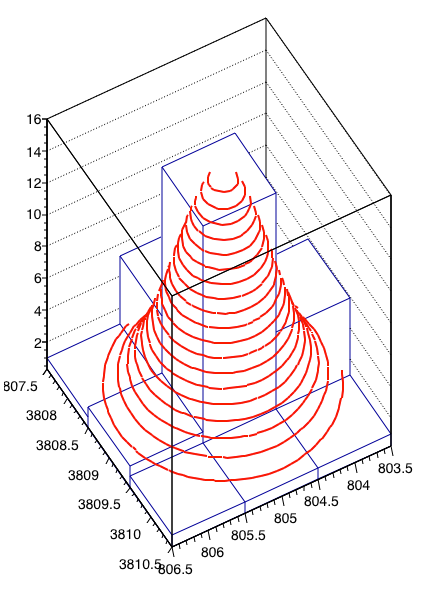}
\caption{Example of a likelihood fit for a 20e$^-$ event.}
\label{fig:likelihoodfit}
\end{center}

\end{figure}

The signature for a nuclear recoil in a CCD is a diffusion-limited hit, as in Fig. \ref{fig:partid}.
These hits are characterized by a two-dimensional Gaussian charge distribution with a width (RMS) determined
by the diffusion of charge inside the CCD, perpendicular to the
drift direction. The diffusion of charge in the CCD depends on the drift distance between the 
location of charge generation and the electrical potential minimum next to the CCD gates, where the charge
is collected. For charge generated at the back of the sensor, on the surface far away from the gates,
the diffusion is approximately 0.5 pixels (7.5 $\mu$m).  Diffusion-limited hits are selected as 
clusters of charge with at least one pixel having more than 10 e$^-$, corresponding to 5 $\sigma$ for the noise
shown in Fig. \ref{fig:noisestab}.   A 2-D Gaussian is adjusted to every hit found in 
the CCD image using a maximum likelihood technique, as shown in Fig. \ref{fig:likelihoodfit}. The likelihood for
the hit being the result of a noise fluctuation is also calculated assuming Gaussian readout noise.

Low-energy electron recoils also produce diffusion-limited hits, these events could
originate from low energy X-rays coming from outside the sensors. Such low-energy X-rays
do not penetrate deeply into the CCD, at 5 keV the attenuation for a photon in silicon is $\sim$
10 $\mu$m. X-rays  converting at the front (back) surface of the sensor will generate lower (higher) diffusion compared
to events occurring in the bulk of the CCD. A selection cut resulting from the likelihood fit is applied on the size
of the hit to suppress the background from low-energy 
X-rays external to the sensor.

\begin{figure}
\begin{center}
\includegraphics[width=0.8\textwidth]{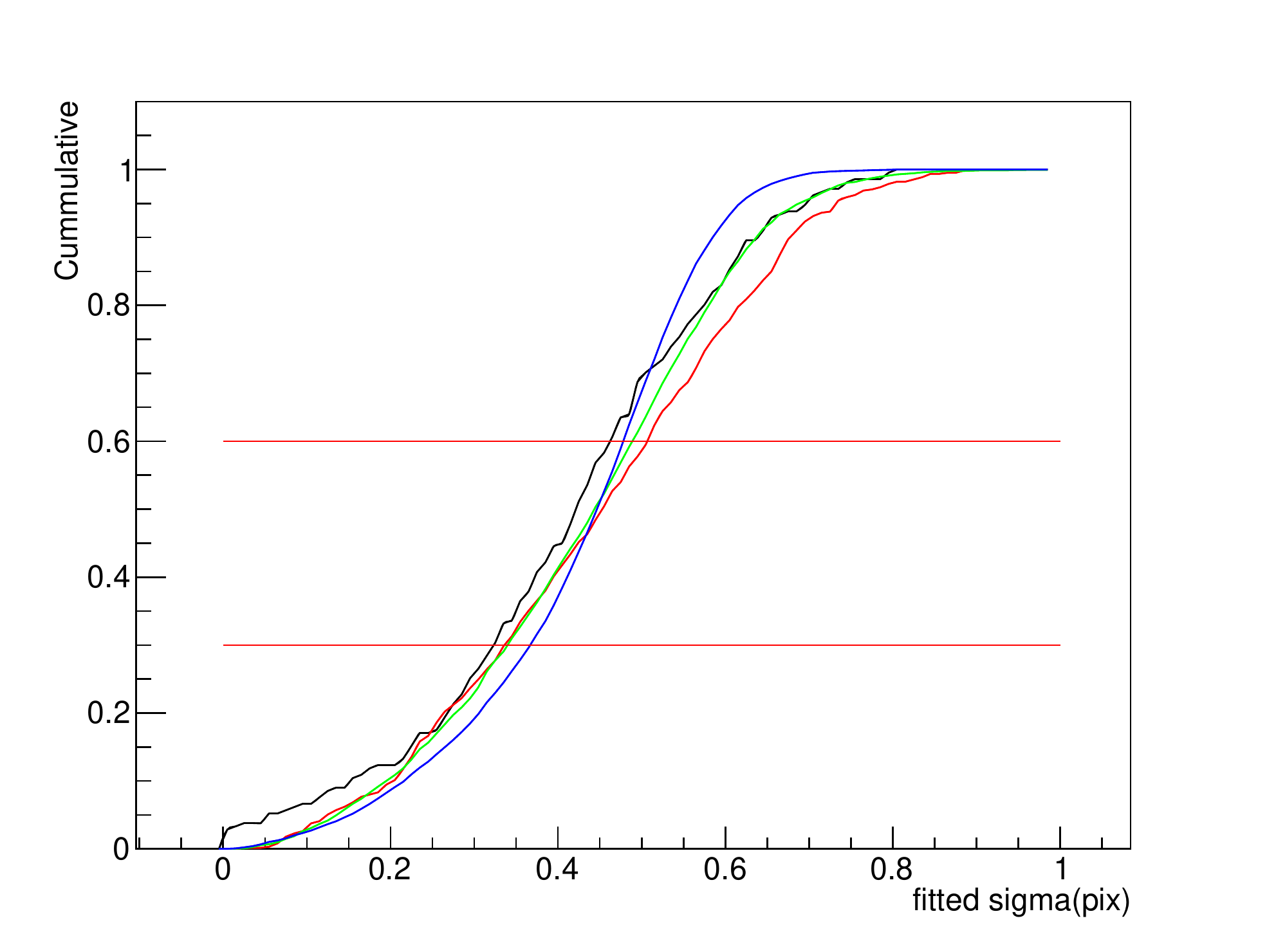}
\caption{Cumulative probability for the fitted size of the reconstructed 2-D hit in simulations, as shown in Fig. \ref{fig:likelihoodfit}. Black, red, green, and blue correspond to $Q < 15$ e- , $15 <Q < 45$ e$^-$, $45 <Q < 135$ e$^-$, and  $135 <Q < 500$ e$^-$. The magnitude of the diffusion for an event depends on the drift distance, and is reflected in the reconstructed size of the hit.}
\label{fig:bulksel}
\end{center}
\end{figure}

\begin{figure}
\begin{center}
\includegraphics[width=0.8\textwidth]{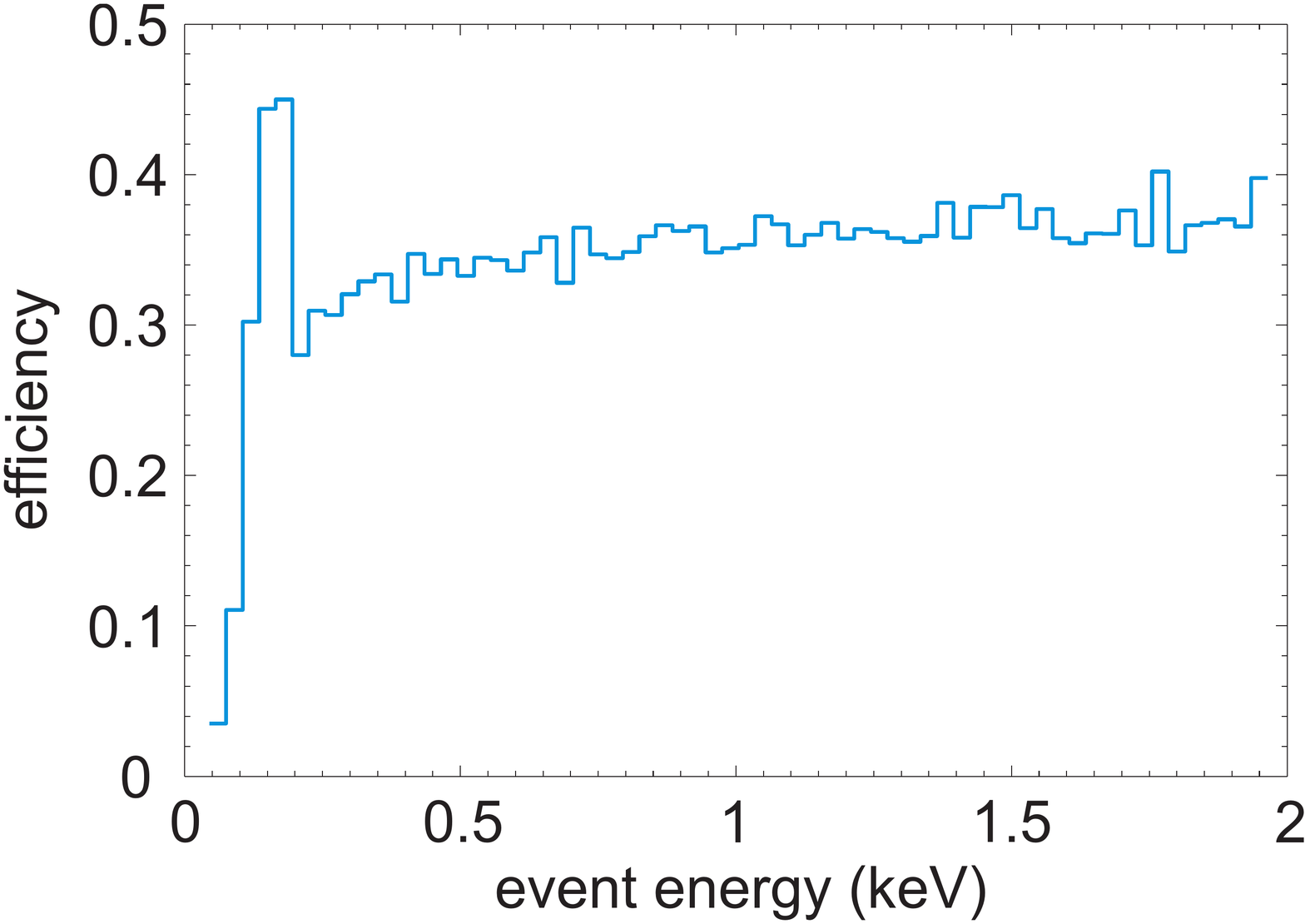}
\caption{Efficiency of the selection criteria discussed in Table \ref{tab:select} measured in simulated events.}
\label{fig:seleff}
\end{center}
\end{figure}

A sample of simulated nuclear-recoil hits, uniformly distributed within the volume of the CCD, 
was generated to study the efficiency of the selection criteria. The cumulative distribution function for
the size of the charge distribution ($\sigma$) in the simulated events is shown in Fig. \ref{fig:bulksel}.  Events occurring 
close to the surface of the sensor are eliminated by selecting a diffusion band as discussed in Table \ref{tab:select}. 
This is approximately equivalent to eliminating a layer of 50 $\mu$m on each surface of the detector.
The efficiency of these selection cuts as a function of energy is shown in Fig. \ref{fig:seleff}.

\begin{table}
\caption{Event selection criteria for three energy regions considered for the analysis. $Q$ is the
total reconstructed charge in the event and $\sigma_D$ the lateral charge diffusion determined
by the 2-D Gaussian fit to each hit. In addition to the selection of $\sigma_D$ and $Q$, the events
are required to have a good quality fit to the 2-D Gaussian hypothesis.}\label{tab:select}
\begin{center}
  \begin{tabular}{ || c |  c   || }
    \hline
    Low Energy  &  $Q <$ 50 e$^-$          				        \\
		       &  $0.12 < \sigma_D <  0.45$ 			\\ \hline \hline
    Medium Energy      &   50 e$^-$ $< Q  <$ 100 e$^-$           				        \\
		       &  $0.23 < \sigma_D <  0.45$ 			\\ \hline \hline
    High Energy                   &   $ Q  >$ 100 e$^-$           				        \\
		       &  $0.28 < \sigma_D <  0.43$ 			\\ \hline \hline
  \end{tabular}
\end{center}
\end{table}

\section{Reactor on/off comparison}

\begin{figure}
\begin{center}
\includegraphics[width=0.8\textwidth]{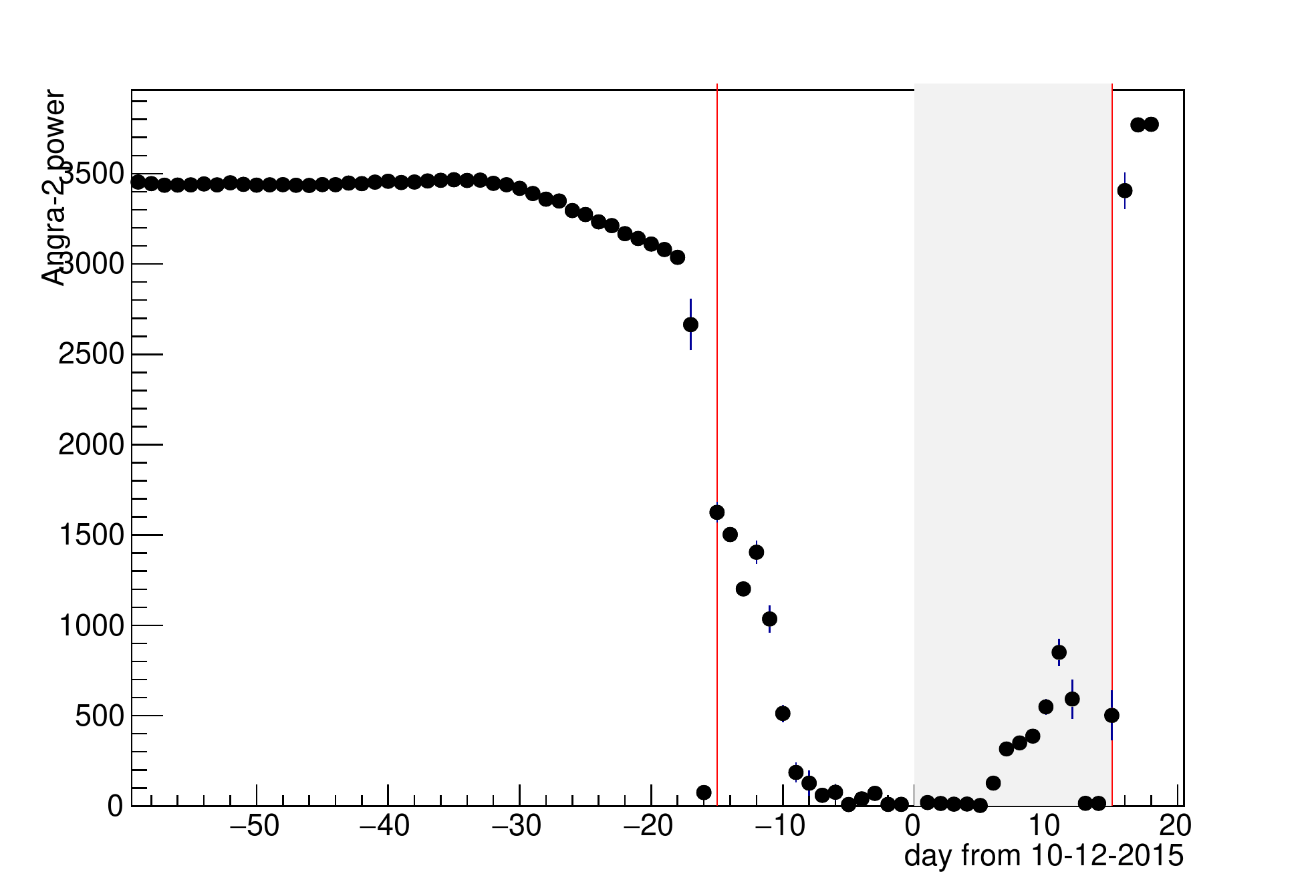}
\caption{Thermal power of the Angra-2 nuclear power plant as a function of time in days.
The zero in the z-axis begins 12-OCT-2015, the start of the low-noise run during
the reactor shutdown. The red vertical lines indicate the reactor shutdown period. The
grey box represented Run-I as shown in Table \ref{tab:runs}.}
\label{fig:AngraPower}
\end{center}
\end{figure}
\begin{figure}
\begin{center}
\includegraphics[width=0.8\textwidth]{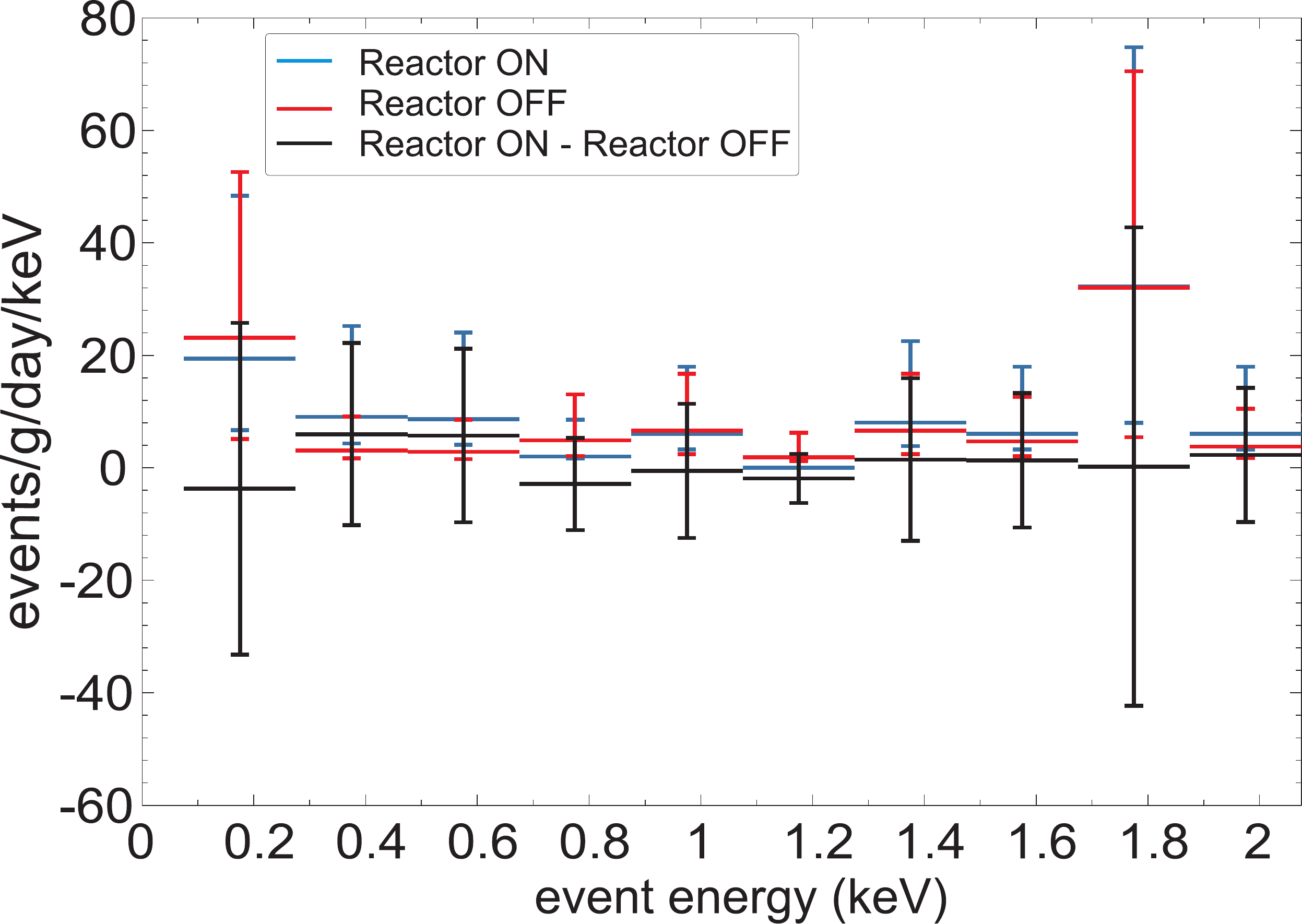}
\caption{Energy spectrum measured with Reactor on, Reactor off and their difference. Events
are selected as discussed in the text, and the rate is corrected for the efficiency of the
selection criteria. The error bars are calculated as the square root of the number of events
in each bin. The higher rate of events at 1.8 keV is
produced by the silicon fluorescence X-ray.}
\label{fig:OnOffComp}
\end{center}
\end{figure}
\begin{figure}
\begin{center}
\includegraphics[width=0.8\textwidth]{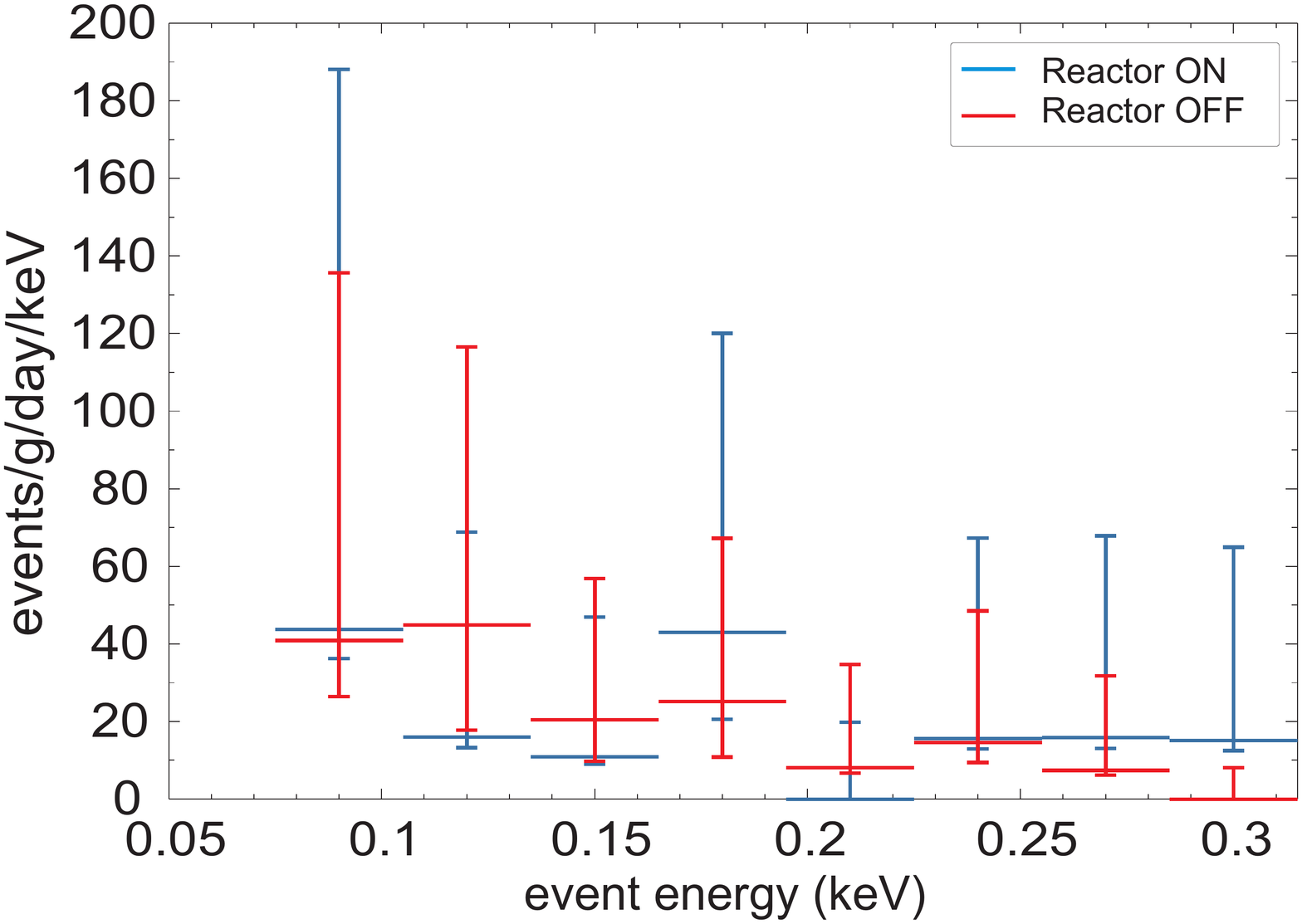}
\caption{Same as Fig.\ref{fig:OnOffComp}, the error bars correspond to 68.27\% probability assuming a
Poisson distribution for each energy bin..}
\label{fig:OnOffComp_lowE}
\end{center}
\end{figure}

As commonly done for this type of reactor there is a programmed shutdown approximately every year, with a typical duration of 
one month. During 2015 this shutdown occurred in October as shown in Fig. \ref{fig:AngraPower}. Data were collected with
the CONNIE engineering array during the normal reactor operations (RON) and during the reactor shutdown (ROFF). Data from the 
first half of the reactor shutdown suffered from a factor of two extra noise in the readout system. 
This noise was caused by the installation of additional equipment operating near the detector, 
related to reactor-shutdown activities and was not part of the CONNIE experiment.  The noise was eliminated by decoupling the CONNIE DAQ from the rest
of the electronic equipment operating on-site. The noisy data during the first half of the shutdown are not used for the analysis discussed here.  
The runs considered for further analysis are summarized in Table \ref{tab:runs}. 
Using the selection criteria discussed above (see Fig. \ref{fig:seleff}), the energy spectra for ROFF/RON periods are compared
in Fig. \ref{fig:OnOffComp} and Fig. \ref{fig:OnOffComp_lowE}.

\begin{table}
\caption{Runs }\label{tab:runs}
\begin{center}
  \begin{tabular}{ || c |  c | c | c | r || }
    \hline
    Run & Reactor &  noise   CCD-A         &  exposure  &  start \\
             &             &   e- RMS       &  day            &   \\ \hline \hline
     I        & off  &  1.8                 & 14.8           &   12-OCT-2015  \\ \hline
     II        & on   &  1.8	           & 7.6             &   27-OCT-2015\\  \hline
     III        & on   &  2.		   & 11.2           &   03-NOV-2015  \\  \hline
     \hline
  \end{tabular}
\end{center}
\end{table}

\section{Conclusion and Prospects}

The CONNIE Collaboration has successfully demonstrated the feasibility of remote operation of a 
CCD array at the Angra-2 nuclear power plant in the state of Rio de Janerio, Brazil. After a few initial
months of commissioning, and establishing optimal remote operating procedures for the detector,
90$\%$ running efficiency was achieved between August 2015 and January 2016. The engineering detector array operated as expected, with noise
around 2e$^-$ and limited by the cosmetic quality of the sensors.  The stability of the noise was demonstrated
when the system was isolated from other equipment running on site.

The radiation background observed in the engineering array is ~3000 events/kg/day/keV (d.r.u.) at 0.5 keV.
This background rate is consistent with radiation coming from the packaging components used
for the engineering array. In particular, the aluminum nitrade (AlN)  board used as mechanical support on the packages
is known to have activity from U and Th producing similar levels of background \cite{Tiffenberg 2013, Damic 2014}. This
material will be eliminated for the next version of the detector packaging. During the engineering run, 
the stability of the gamma background was demonstrated to better than 3\% using the fluorescence peaks from Cu.

The good quality data collected in the engineering run included 15 days during the Angra-2 annual reactor shutdown. This
sample allowed a comparison of reactor on/off to establish a limit for the event
rate coming from neutrinos produced at the reactor. The limit is approximately 2 orders of magnitude
above the rate expected for the standard model neutrino nucleus coherent scattering. These results
start probing an interesting region of new physics in the low energy neutrino sector as discussed in Ref \cite{Harnik 2012}. 

The CONNIE Collaboration is now planning an upgrade to $\sim$100 g of detector during 2016. The plan also includes an
upgrade in the packaging design to reduce the internal background of the detector (elimination of AlN parts). The upgrade
will incorporate engineering grade sensors with 675 $\mu$m thickness and 5.7~g of active mass per sensor, for a total of 18 CCDs. The
upgraded detector should achieve sensitivity to the standard model neutrino-nucleus coherent scattering as discussed in
\cite{Moroni 2015}.

\section*{Acknowledgments}

We thank Central Nuclear Almirante \'Alvaro Alberto Eletronuclear, for access to the Angra II reactor site and for the support of their personnel, in particular Ilson Soares, to the CONNIE activities. We thank the Silicon Detector Facility team at Fermi National Accelerator Laboratory for being the host lab for the assembly and testing of the detectors components used in the CONNIE experiment. We acknowledge the support from Minist\'erio da Ciencia, Tecnologia e Inovacao (MCTI) and the Brazilian funding agencies FAPERJ (grant E-26/110.145/2013), CNPq, and FINEP; and the Mexiacan agencies PAPIIT-UNAM (grants IB100413, and IN112213), and CONACYT (grant No. 240666); and Agencia Nacional de Promoci\'on Cient\'ifica y Tecnológica in Argentina. We thank Ricardo Galvão and Ronald Shellard for their support to the experiment.


\section*{References}

\begin{thebibliography}{99}


\bibitem{Hasert 1973}
Hasert F. J. et al., Phys. Lett B, 46 (1973) 121 and 138.

\bibitem{Freedman 1974}
Freedman D. Z., Coherent effects of a weak neutral current, Phys. Rev. D 9, 1389-1392 (1974)

\bibitem{Freedman 1977}
Freedman D. Z., Schramm D. N. and Tubbs D. L., Annu. Rev. Nucl. Part. Sci., 27 (1977) 167.

\bibitem{Scholberg 2006}
Scholberg K.,  Phys. Rev. D 73, 033005 (2006).


\bibitem{Wilson 1974}
Wilson J.R.,  Phys. Rev. Lett. 32, 849 (1974).

\bibitem{Horowitz 2004} 
Horowitz C.J., Perez-Garcia M.A,Carriere J., Berry D.K. , and Piekarewicz J.. Phys. Rev. C70, 065806 (2004). [astro-ph/0409296].

\bibitem{Barbeau 2003}
Barbeau et al. IEEE Trans. Nucl. Sci. 50: 1285 (2003).

\bibitem{Hagmann 2004}
Hagmann C. and A. Bemstein, IEEE Trans. Nucl. Sci 51:2151 (2004).

\bibitem{Snowmass 2013}
Cushman P., et al. Snowmass CF1 Summary: WIMP Dark Matter Direct Detection (2013)  [arXiv:1310.8327] 


\bibitem{Billard 2014} J.Billard et al, Phys. Rev. D 89, 023524 (2014), [arxiv:1307.5458]

\bibitem{Strigari 2009} L.E.Strigari   (2009) [arxiv:0903.3630]

\bibitem{Glutein 2010} A.Glutein et al, (2010) [arxiv:1003.5530]

\bibitem{Harnik 2012} R. Harnik et al, JCAP 07 (2012) 026, [arxiv:1202.6073]

\bibitem{Anderson 2012} A. J. Anderson, J. M. Conrad, E. Figueroa-Feliciano, C. Ignarra, G. Karagiorgi, K. Scholberg, M. H. Shaevitz, and J. Spitz,  Phys. Rev. D 86, 013004 (2012).

\bibitem{Dutta 2015} B. Dutta et al. (2015)  [arXiv:1511.02834]

\bibitem{doublechooz}   M. Apollonio et al. [CHOOZ Collaboration], Phys. Lett. B 466 (1999) 415-430

\bibitem{Boehm 2000}
Boehm F., Phys. Rev. D 64, 112001(2001).

\bibitem{kamland} K. Eguchi et al. (KamLAND Collaboration), Phys. Rev. Lett. 90, 021802, 2003.


\bibitem{Zacek} G. Zacek et al. [CALTECH-SIN-TUM Collaboration] , Phys. Rev. D 34, 2621 (1986).

\bibitem{Vidyakin} G. S. Vidyakin et al., JETP Lett. 59 (1994) 390 [Pisma Zh. Eksp. Teor. Fiz. 59, 364 (1994)].

\bibitem{Declais} Y. Declais et al., Nucl. Phys. B 434, 503 (1995).

\bibitem{An} F. P. An, J. Z. Bai, A. B. Balantekin, et al.,  Physical Review
Letters, vol. 108, no. 17, Article ID 171803, 2012.

\bibitem{Ahn} J. K. Ahn, S. Chebotaryov, J. H. Choi, et al., Physical Review Letters, vol. 108, no. 19, Article ID 191802, 6 pages, 2012.

\bibitem{Abe} Y. Abe, C. Aberle, J. C. dos Anjos, et al., Physics Letters B, vol. 723, no. 1?3, pp. 66?70, 2012.


\bibitem{Xin 2005}
Xin B., Production of electron neutrinos at nuclear power reactors and the prospects for neutrino physics, Phys. Rev. D 72, 012006 (2005).

\bibitem{GEMMA} A.G. Beda et al., Phys. Part. Nucl. Lett. 10 139-143 (2013).

\bibitem{Wong 2007}
Wong, H. T. et al, Search of neutrino magnetic moments with a high-purity germanium detector at the Kuo-Sheng nuclear power station, Phys. Rev. D 75, 012001 (2007).

\bibitem{Coherent 2015} COHERENT Collaboration, The COHERENT Experiment at the Spallation Neutron Source, 1509.08702 (2015)

\bibitem{Boyle 2010}
Boyle, W.S.: "Nobel Lecture: CCD---An extension of man's view" ,Rev. Mod. Phys.,
\textbf{82}(3), 2305--2306, (2010)

\bibitem{Smith 2010}
Smith, G.E.: Nobel Lecture: The invention and early history of the CCD, Rev. Mod. Phys. \textbf{82} (3), 2307--2312, (2010)

\bibitem{Janesick 2001}
Janesick, J.R.: {S}cientific {C}harge {C}oupled {D}evices, {SPIE}
  {P}ublications, Bellingham, Washington, (2001)


\bibitem{Holland 2003}
Holland, S.E. et al.: Fully depleted, back-illuminated charge-coupled devices fabricated on high-resistivity silicon, {IEEE} Trans. Electron Devices \textbf{50}(1), 225--238  (2003)


\bibitem{Barreto 2012}
Barreto J. et al, Direct search for low mass dark matter particles with CCDs, Physics Letters B 711, 264-269 (2012).

\bibitem{Tiffenberg 2013}
Tiffenberg, J. S., \label{tab:runs}. ICRC 2013.

\bibitem{Aguilar 2015}
Aguilar-Arevalo A. et al, Measurement of radioactive contamination in the high-resistivity silicon CCDs of the DAMIC experiment.  arXiv:1506.02562 JINST 10 (2015) no.08, P08014

\bibitem{Antonella_cal} J. Tiffenberg, "Results from the Antonella experiment", https://kicp-workshops.uchicago.edu/2015-lowecal/depot/talk-tiffenberg-javier.pdf (2015)

\bibitem{Uchicago_cal} A. Chavarria, "Response of a CCD to low energy nuclear recoils from a 124Sb-9Be photo-neutron source", https://kicp-workshops.uchicago.edu/2015-lowecal/depot/talk-chavarria-alvaro.pdf (2015)


\bibitem{Lindhard 1963}
J. Lindhard, V. Nielsen, M. Scharff, P.V. Thomsen, Mat. Fys. Medd. Dan. Selsk. 33 (1963) 10.

\bibitem{Lewin 1996}
J.D. Lewin, P.F. Smith, Astropart. Phys. 6 (1996) 87.

\bibitem{cryomech} www.cryomech.com

\bibitem{lakeshore} www.lakeshore.com

\bibitem{Oluseyo 2004}  H. Oluseyo et al, Sensors, Systems, and Next-Generation Satellites VIII. Edited by R. Meynart, S. P. Neeck, H.
Shimoda, SPIE 5570, 515-524 (2004).

\bibitem{iaea_pris} https://www.iaea.org/pris/

\bibitem{Moroni 2015}  
Fernandez-Moroni G. et al., "Charge Coupled Devices for detection of coherent neutrino-nucleus scattering", Phys. Rev. D 91, 072001 (2015)

\bibitem{conversionFactor}  D. Groom et al,  "Temperature dependence of mean number of of e-h pairs per eV of x-ray energy deposit", www-ccd.lbl.gov/w$\_$Si.pdf  (2004)

\bibitem{Damic 2014}  Chavarria A.E. et al. Physics Procedia, Volume 61, p. 21-33.(2014) arXiv:1407.0347


\end{thebibliography}
\end{document}